\documentclass[preprint,aps,prd,superscriptaddress,nofootinbib]{revtex4-2}
\usepackage{amsmath,amssymb,mathrsfs}
\usepackage[normalem]{ulem}
\usepackage[colorlinks=true,linkcolor=blue,citecolor=blue,urlcolor=blue]{hyperref}

\newcommand{\dd}{\mathrm{d}}
\newcommand{\slasheddelta}{\not\!\delta}
\newcommand{\Omegac}{\Omega_{\mathrm c}}

\begin{document}

\title{Thermodynamics of Kerr--Newman--Bertotti--Robinson black holes}

\author{Zelin Zhang}
\affiliation{Institute of Fundamental Physics and Quantum Technology, \& School of Physical Science and Technology, Ningbo University, Ningbo, Zhejiang 315211, P. R. China}

\author{Zhenyu Zhang}
\affiliation{Institute of Fundamental Physics and Quantum Technology, \& School of Physical Science and Technology, Ningbo University, Ningbo, Zhejiang 315211, P. R. China}

\author{Bin Chen}
\email{Corresponding author: chenbin1@nbu.edu.cn}
\affiliation{Institute of Fundamental Physics and Quantum Technology, \& School of Physical Science and Technology, Ningbo University, Ningbo, Zhejiang 315211, P. R. China}
\affiliation{School of Physics, \& Center for High Energy Physics, Peking University, No.5 Yiheyuan Rd, Beijing 100871, P. R. China}

\begin{abstract}
In this work, we extend the thermodynamic analyses of the neutral and specially charged
Kerr--Bertotti--Robinson black holes to the general
Kerr--Newman--Bertotti--Robinson family, in which the electric and
external-field parameters are independent.  Using covariant surface charges and canonical integrability
methods, we determine the total angular momentum and the canonical mass.
The angular momentum follows analytically from the combined
gravitational and electromagnetic surface charges. However, the infinitesimal
charge associated with coordinate time translations is not integrable in
solution space, so the mass must be associated with a more general symmetry
generator.
Imposing the canonical integrability conditions on this generator, together
with the Kerr--Newman mass as the zero-field boundary condition, selects the
Christodoulou--Ruffini mass and determines the associated thermodynamic
potentials.  The first law and Smarr formula take the same form as the ones in usual Kerr--Newman case, and the two previously studied Kerr--Bertotti--Robinson cases are recovered as special limits. 
\end{abstract}

\maketitle
\newpage
\section{Introduction}
\label{sec:introduction}

The laws of black-hole mechanics, first formulated by Bardeen, Carter, and
Hawking for stationary solutions \cite{BardeenCarterHawking}, acquired
their thermodynamic interpretation through Bekenstein's entropy--area
relation and Hawking radiation \cite{Bekenstein1973,Hawking1975}.  In the
covariant phase-space formulation, the first law is expressed in terms of
the Hamiltonian charges associated with the spacetime symmetries, and the
entropy is identified with a horizon Noether charge
\cite{LeeWald,Wald1993,IyerWald}.
A complete thermodynamic description therefore requires consistent
definitions of the mass, angular momentum, electric charge, and their
conjugate potentials.  For asymptotically flat stationary black holes,
these definitions are supplied by the asymptotic structure.  The mass and
angular momentum are generated by the asymptotic Killing fields, and the
electrostatic potential is defined relative to infinity.

These choices are no longer supplied automatically when a black hole is
embedded in an external field and the spacetime is not asymptotically flat.
The total angular momentum may then contain an
electromagnetic contribution \cite{AshtekarRotatingIH}, and its value
depends on choosing a regular gauge potential and a definite axial
generator.  Although the horizon still satisfies a local mechanical
identity, the infinitesimal charge associated with a coordinate time
translation need not be integrable in solution space.  The mass must then
be defined together with the symmetry parameter that generates it.
Exact magnetized black holes provide an important setting for this
problem.  The Ernst--Wild solution describes a Kerr--Newman black hole
embedded in a backreacting magnetic field and approaches the Melvin
magnetic universe in the asymptotic region
\cite{ErnstWild1976,Melvin1964,GibbonsMagnetized}.  Several prescriptions
for its conserved charges and thermodynamics have been investigated
\cite{GibbonsPangPope,BoothMelvin}.

Astorino et al.\ applied the canonical integrability methods developed in
the surface-charge formalism \cite{BarnichBrandt,BarnichBoundary,
BarnichCompere} to this Kerr-Newman-Melvin geometry \cite{AstorinoCompere}.  Their construction
starts from the infinitesimal surface charge associated with a general
symmetry parameter and defines the mass by imposing the integrability
conditions under variations of all solution parameters, including the
external field. For the Kerr--Newman--Melvin 
family, the zero-field Kerr--Newman mass provides the boundary condition
that selects a canonical mass obeying the Christodoulou--Ruffini relation
\cite{ChristodoulouRuffini,AstorinoCompere}.  Once the canonical generator
is known, a change of coordinates and gauge defines the canonical frame in
which this mass is the charge of time translations.  The same mass agrees
with the quasi-local energy obtained from the rotating isolated-horizon
framework \cite{AshtekarRotatingIH,BoothMelvin}. 
The same surface-charge framework, together with integrability in solution
space, has been applied to black holes with unusual asymptotics
\cite{BarnichCompereGodel,BanadosGodel}.

The Bertotti--Robinson universe offers a different exact electromagnetic
background.  It is the homogeneous $\mathrm{AdS}_2\times S^2$ solution of
the Einstein--Maxwell equations supported by a uniform electromagnetic field
\cite{Bertotti1959,Robinson1959}.  A rotating black hole immersed in this
background was recently constructed as an exact solution whose
electromagnetic field is not aligned with the principal null directions
\cite{PodolskyOvcharenko2025,OvcharenkoPodolsky2025}.  It belongs to the
general Kerr--Newman--Bertotti--Robinson (KN--BR) family, whose four
parameters $(m,a,e,B)$ set the mass, rotation, electric-charge, and
external-field
scales but do not all coincide directly with the physical charges
\cite{OPnew}.  The family reduces to Kerr--Newman when the external field is
removed.  It also contains two Kerr--Bertotti--Robinson (Kerr--BR)
restrictions, namely the specially charged Kerr--BR$_s$ branch, on which
the rotating solution was originally discovered, and the genuine neutral
Kerr--BR$_0$ branch, with the subscript notation following
Ref.~\cite{OPnew}.
This class of exact rotating geometries in the Bertotti--Robinson universe has already motivated intense studies from various points of view, ranging from hidden
symmetries and separability, null geodesics and shadows, circular orbits and
inspirals, the Kerr/CFT correspondence, wave signatures, dynamical
properties, to global structure of spacetime
\cite{GrayEtAl2026,WangISCO2025,liu2026,SiahaanKerrCFT2026,
LiGW2026,MustafaEtAl2026,WangShadow,Wan:2026lca,Wan:2026bkb,ZhouEtAl2026,Zhou:2026fvj}.  Its thermodynamics has so far been
worked out only for the two Kerr--BR restrictions.  Hu et al. determined
the conserved charges of the specially charged branch, adopted the
Christodoulou--Ruffini relation as the definition of its mass, and
identified the associated solution-dependent generator \cite{HuCaiWang}.
Kubiz\v{n}\'ak et al. established the thermodynamics of the neutral
branch and showed that, after an appropriate rescaling of the coordinates
and parameters, its thermodynamic relations take the Kerr form
\cite{KOPthermo}.

In this paper, we investigate the thermodynamics of the general KN--BR
family using canonical integrability methods. Our study is different from the specially charged case Kerr--BR$_s$, in which the electric parameter is constrained by the remaining parameters.  We first adopt the regular
conventions for the axial coordinate and the Maxwell potential required
for the surface charges.  We then compute the electric charge and the
total angular momentum and examine the integrability of the infinitesimal
charge associated with the coordinate time translation.  The mass is then
determined by a general canonical generator subject to the zero-field
Kerr--Newman boundary condition.  We subsequently determine the
thermodynamic potentials, the canonical frame, and the relation to the two
previously studied Kerr--BR cases. 

The paper is organized as follows.  Section~\ref{sec:geometry} introduces
the general KN--BR solution and its horizon quantities.  Section~\ref{sec:charges}
reviews the canonical surface-charge construction, evaluates the electric
charge and angular momentum, and identifies the mass problem.  The canonical
mass is determined from integrability in Sec.~\ref{sec:mass}.  Its
thermodynamic properties, canonical frame, alternative fixed-source
description, and Kerr--BR limits are discussed in Sec.~\ref{sec:thermo}. Some technical details are provided in the appendices. We use units $G=c=\hbar=k_B=1$.

\section{The general Kerr--Newman--Bertotti--Robinson black hole}
\label{sec:geometry}

\subsection{Spacetime geometry and regularity conditions}

The general KN--BR solution was obtained in Ref.~\cite{OPnew}.  Following
the convention of a $2\pi$-periodic azimuthal angle used in the
thermodynamic analyses of the Kerr--BR subfamilies \cite{HuCaiWang,KOPthermo},
we employ coordinates $(t,r,\theta,\phi)$ adapted to the stationary and axial
symmetries.  The metric is
\begin{align}
 \dd s^2=\frac{1}{\Omegac^2}\bigg[
 -\frac{Q}{\rho^2}
 \left(\dd t-\frac{a\sin^2\theta}{K}\dd\phi\right)^2
 +\frac{\rho^2}{Q}\dd r^2+\frac{\rho^2}{P}\dd\theta^2+\frac{P\sin^2\theta}{\rho^2}
 \left(a\dd t-\frac{r^2+a^2}{K}\dd\phi\right)^2\bigg],
 \label{eq:metric}
\end{align}
where
\begin{align}
 \rho^2&=r^2+a^2\cos^2\theta,
 \quad P=1+B^2\mu^2\cos^2\theta,
 \quad Q=I\Delta,
 \nonumber\\
 I&=(1+kBr)^2+B^2r^2,
 \qquad \Omegac^2=I-B^2\Delta\cos^2\theta,
 \nonumber\\
 \Delta&=(1+k^2)a^2-2mr +\left(1+k^2-\frac{e^2}{a^2}\right)r^2.
 \label{eq:metric-functions}
\end{align}
The auxiliary constants are
\begin{align}
 s^2&=1+(e^2-a^2)B^2-(m^2+e^2)a^2B^4,
 \quad k=\frac{es-amB}{a(1+e^2B^2)},
 \nonumber\\
 \mu^2&=m^2-(1+k^2)^2a^2+(1+k^2)e^2,
 \qquad K\equiv1+B^2\mu^2.
 \label{eq:auxiliary-functions}
\end{align}
Here $\Omegac$ denotes the conformal factor called $\Omega$ in
Ref.~\cite{OPnew}.  The subscript distinguishes it from the thermodynamic
angular velocity introduced below. The factor $K$ is the inverse of the conicity parameter $C$ of
Ref.~\cite{OPnew}.  In the original coordinates the regular angular
identification has period $2\pi C=2\pi/K$.  The factors of $K^{-1}$ in
Eq.~\eqref{eq:metric} absorb this conicity, so that $\phi$ retains the
standard period $2\pi$.  The axial Killing vector used throughout the paper
is therefore
\begin{equation}
 \eta=\partial_\phi .
 \label{eq:axial-Killing}
\end{equation}

In this work, we take the convention that both $\mu$ and $s$ are taken to be positive, $\mu=+\sqrt{\mu^2}, s=+\sqrt{s^2}$. The roots of $\Delta$, giving  the horizons, are ordered as
\begin{equation}
 r_\pm=\frac{m\pm\mu}{1+k^2-e^2/a^2},
 \qquad r_+\geq r_-.
 \label{eq:horizon-radii}
\end{equation}
We restrict to parameter values for which $r_+$ is a regular outer horizon
and the exterior has Lorentzian signature.  At fixed nonzero $e$ and $B$, the general
charged parametrization is singular as $a\to0$.  The regular nonrotating
limits are obtained either by setting $e=0$ before taking $a\to0$, which
gives Schwarzschild--BR, or by setting $B=0$ first, which gives
Reissner--Nordstr\"om \cite{OPnew}.

Unlike the metric, which involves only $\Omegac^2$, the Maxwell
potential contains $\Omegac^{-1}$.  We therefore choose the branch of
$\Omegac$ connected continuously to $\Omegac=1$ in the zero-field
Kerr--Newman limit, and restrict to the exterior region where it remains
positive.

The $2\pi$-periodic axial generator must be paired with a Maxwell potential
that is regular on both symmetry axes.  This is required because the total
angular momentum contains an electromagnetic surface term involving
$\eta\cdot A$.  Let $A^{(0)}$ denote the potential of Ref.~\cite{OPnew} with the arbitrary
closed one-form set to zero, re-expressed in terms of the azimuthal
coordinate $\phi$.  Its value on either axis is $A^{(0)}_\phi=-u/(BK)$, where
\begin{equation}
 u=\frac{s+ameB^3}{1+e^2B^2}.
 \label{eq:u-definition}
\end{equation}
The regular representative is therefore
\begin{equation}
 A=A^{(0)}+\frac{u}{BK}\,\dd\phi,
 \qquad A_\phi\big|_{\theta=0,\pi}=0.
 \label{eq:axis-gauge}
\end{equation}
The equal values at the two poles also imply a vanishing magnetic monopole
charge.  The explicit components are recorded in Appendix~\ref{app:charges}.
Expressions containing $1/B$ are understood through their smooth zero-field
limit, which reproduces the standard Kerr--Newman potential.  With the axial
generator and the Maxwell representative fixed, the surface charges can be
compared consistently across the solution family.

\subsection{Horizon quantities}

The outer horizon is generated by
\begin{equation}
 \chi=\partial_t+\Omega_H\partial_\phi.
 \label{eq:horizon-generator}
\end{equation}
It is convenient to introduce the area radius of the horizon
\begin{equation}
 R_H^2\equiv\frac{A_H}{4\pi}=\frac{S}{\pi}
 =\frac{r_+^2+a^2}{K I_+},
 \qquad I_+\equiv I(r_+).
 \label{eq:area-radius}
\end{equation}
The entropy and the Hawking temperature associated with the horizon
generator $\chi$ are
\cite{OPnew}
\begin{equation}
 S=\pi R_H^2,
 \qquad T_H=\frac{\mu}{2\pi K R_H^2}.
 \label{eq:entropy-temperature}
\end{equation}
The horizon angular velocity with respect to $\partial_\phi$ is
\begin{equation}
 \Omega_H=\frac{aK}{r_+^2+a^2}.
 \label{eq:horizon-angular-velocity}
\end{equation}
In the axis-regular gauge, the electrostatic potential is constant over the
horizon and takes the form
\begin{equation}
 \Phi_H=-\chi\cdot A\big|_{r_+}
 =\frac{er_+}{r_+^2+a^2}.
 \label{eq:horizon-potential}
\end{equation}
The external field enters this expression through $r_+$.  The horizon
quantities also satisfy
\begin{equation}
 2T_HS=\frac{\mu}{K}.
 \label{eq:TS-identity}
\end{equation}
These are the local quantities associated with the Killing generator
$\chi$.  Their relation to the thermodynamic potentials will be fixed by the
canonical generator of the mass.

\section{Canonical surface charges and the mass problem}
\label{sec:charges}

\subsection{Canonical integrability methods}

In Einstein--Maxwell theory, symmetry parameters are composed of a pair
$\epsilon=(\xi,\lambda)$ satisfying the generalized Killing equations
\cite{BarnichBrandt,BarnichBoundary,BarnichCompere,CompereReview}
\begin{equation}
 \mathcal L_\xi g_{\mu\nu}=0,
 \qquad \mathcal L_\xi A+\dd\lambda=0.
 \label{eq:generalized-Killing}
\end{equation}
Here $\xi$ is a Killing vector.  The Maxwell potential is chosen to be invariant
under the symmetries considered below, so that $\lambda$ is a real
constant.  The surface-charge form
$\boldsymbol{k}_\epsilon[\delta g,\delta A;g,A]$ is a two-form in spacetime
and a one-form in field space.  Boldface distinguishes this form from the
metric parameter $k$ introduced in Eq.~\eqref{eq:auxiliary-functions}.  For fixed symmetry parameters, the integral of the surface charge form over a closed two-surface
$\Sigma$ defines the infinitesimal charge
\begin{equation}
 \delta\mathcal Q_\epsilon
 =\int_{\Sigma}\boldsymbol{k}_\epsilon
 [\delta g,\delta A;g,A].
 \label{eq:surface-charge}
\end{equation}
For an exact symmetry and a linearized solution,
$\boldsymbol{k}_\epsilon$ is closed on shell.  The integral is therefore
unchanged under deformations of $\Sigma$ through the source-free
exterior \cite{WaldZoupas,BarnichBrandt,BarnichBoundary}.

A finite charge is obtained by integrating
Eq.~\eqref{eq:surface-charge} along a path in the space of solutions.  Path
independence requires the integrability condition
\begin{equation}
 \int_{\Sigma}\!\left(
 \delta_1\boldsymbol{k}_\epsilon[\delta_2 g,\delta_2 A;g,A]
 -\delta_2\boldsymbol{k}_\epsilon[\delta_1 g,\delta_1 A;g,A]
 \right)=0.
 \label{eq:integrability-condition}
\end{equation}
Following Ref.~\cite{AstorinoCompere}, we refer to this surface-charge
construction and the associated integrability conditions as the canonical
integrability methods.  Independence of the integration surface is a
spacetime conservation statement, whereas integrability ensures that the
finite charge is independent of the path in solution space
\cite{WaldZoupas,BarnichCompere,AstorinoCompere}.

\subsection{Electric charge and total angular momentum}

We choose the orientation of the horizon cross-section toward increasing $r$ and use
$\epsilon_{tr\theta\phi}=+\sqrt{-g}$.  The electric charge and angular
momentum are associated with the symmetry parameters $(0,-1)$ and
$(-\eta,0)$, respectively,
\begin{equation}
 \delta Q_e=\int_{\Sigma}\boldsymbol{k}_{(0,-1)},
 \qquad
 \delta J=\int_{\Sigma}\boldsymbol{k}_{(-\eta,0)}.
 \label{eq:charge-generators}
\end{equation}
For a stationary axisymmetric electrovacuum with a regular Maxwell
potential, the corresponding horizon expressions are
\begin{equation}
 Q_e=\frac{1}{4\pi}\int_{\mathcal H}\!\star F,
 \qquad
 J=\frac{1}{16\pi}\int_{\mathcal H}\!\star\dd\eta^\flat
 +\frac{1}{4\pi}\int_{\mathcal H}\!(\eta\cdot A)\star F,
 \label{eq:horizon-charges}
\end{equation}
where $\eta^\flat=g_{\mu\nu}\eta^\nu\dd x^\mu$ and $\mathcal H$ denotes the
horizon.  The second contribution to
$J$ is the electromagnetic part of the total angular momentum
\cite{AshtekarRotatingIH}.

The electric and axial charge variations satisfy the integrability
condition in Eq.~\eqref{eq:integrability-condition}.  Their angular
integrals can be performed analytically and reduce to boundary terms at
the two axes, as detailed in Appendix~\ref{app:charges}
\begin{equation}
 Q_e=\frac{e}{K},
 \qquad J=\frac{ma}{K^2}.
 \label{eq:physical-charges}
\end{equation}
The electric charge reproduces the result of Ref.~\cite{OPnew}.  The
angular momentum follows from the sum of the gravitational and
electromagnetic contributions.  Its variation is
\begin{equation}
 \delta J=\frac{a\,\delta m+m\,\delta a}{K^2}
 -\frac{2ma}{K^3}\,\delta K,
 \label{eq:variation-J}
\end{equation}
where the last term accounts for the parameter dependence of the regular
angular identification.

\subsection{The coordinate-time charge}

With the entropy, angular momentum, and electric charge determined, we now
turn to the mass.  The first symmetry parameter to examine is the coordinate
time translation $(\partial_t,0)$.  The relevant horizon identity follows
from the on-shell conservation of the surface charge associated with
$\chi$ and the standard Einstein--Maxwell horizon surface-charge identity
\cite{BardeenCarterHawking,IyerWald,RogatkoFirstLaw,GaoFirstLaw}.  The
axis-regular potential and the absence of magnetic monopole charge avoid the
gauge-patch contributions that arise for dipole or monopole charges
\cite{CopseyHorowitz}.  The background value of $\Omega_H$ is held fixed
when the surface-charge form is evaluated
\begin{equation}
 \int_{\mathcal H}\boldsymbol{k}_{(\chi,0)}
 =T_H\delta S+\Phi_H\delta Q_e.
 \label{eq:horizon-charge-identity}
\end{equation}
Using $\chi=\partial_t+\Omega_H\partial_\phi$ and the linearity of the
surface charge gives
\begin{equation}
 \slasheddelta\mathcal Q_{(\partial_t,0)}
 =T_H\delta S+\Omega_H\delta J+\Phi_H\delta Q_e.
 \label{eq:coordinate-time-charge}
\end{equation}
The slash anticipates the nonintegrability demonstrated below and distinguishes
this one-form from the exact variation of a finite charge.

The obstruction already appears on the neutral Kerr--BR$_0$
branch.  Using the notation of Ref.~\cite{KOPthermo}, define
\begin{equation}
 b=\sqrt{1+m^2B^2},
 \qquad K_a=1-a^2b^2B^2,
 \qquad K=b^2K_a.
 \label{eq:neutral-definitions}
\end{equation}
Using the neutral horizon data and $J=ma/K^2$,
Eq.~\eqref{eq:coordinate-time-charge} becomes
\begin{equation}
 \left.\slasheddelta\mathcal Q_{(\partial_t,0)}\right|_{e=0}
 =b\,\delta\!\left(\frac{m}{Kb}\right).
 \label{eq:neutral-time-charge}
\end{equation}
Writing the one-form as $q_m\delta m+q_a\delta a+\cdots$, its curl on a
fixed-$B$ slice is
\begin{equation}
 \left.\left(\frac{\partial q_a}{\partial m}
 -\frac{\partial q_m}{\partial a}\right)\right|_{e=0,B}
 =\frac{2m^2aB^4b^2}{K^2},
 \label{eq:nonintegrability}
\end{equation}
which is nonvanishing for rotating black holes in a nonzero external field.  The same
nonintegrability appears on the specially charged branch
\cite{HuCaiWang}.  Hence $(\partial_t,0)$ is not the symmetry parameter
associated with the thermodynamic energy of the general KN--BR family.

\section{Definition of the mass from integrability}
\label{sec:mass}

\subsection{The canonical generator and first law}

Following the construction of Ref.~\cite{AstorinoCompere}, we consider the
most general symmetry parameter associated with the energy,
\begin{equation}
 \epsilon_M=\alpha
 \bigl(\partial_t+\Omega_{\rm int}\partial_\phi,
 \Phi_{\rm int}\bigr).
 \label{eq:mass-generator}
\end{equation}
The quantities $\alpha$, $\Omega_{\rm int}$, and $\Phi_{\rm int}$ are
constants in spacetime but functions on the solution space.  In the adjusted
surface-charge variation, they are held fixed inside
$\boldsymbol{k}_{\epsilon_M}$, while their parameter dependence is
included when the integrability condition is imposed
\cite{HajianSheikhJabbari,AstorinoCompere}.  By definition,
\begin{align}
 \delta M
 &=\alpha\left[
 \slasheddelta\mathcal Q_{(\partial_t,0)}
 -\Omega_{\rm int}\delta J-\Phi_{\rm int}\delta Q_e\right]
 \nonumber\\
 &=\alpha\left[T_H\delta S
 +(\Omega_H-\Omega_{\rm int})\delta J
 +(\Phi_H-\Phi_{\rm int})\delta Q_e\right].
 \label{eq:mass-variation}
\end{align}
It is useful to introduce the frame-independent thermodynamic potentials
\cite{AstorinoCompere}
\begin{equation}
 T=\alpha T_H,
 \qquad \Omega=\alpha(\Omega_H-\Omega_{\rm int}),
 \qquad \Phi=\alpha(\Phi_H-\Phi_{\rm int}).
 \label{eq:thermodynamic-definitions}
\end{equation}
The canonical first law then takes the standard form
\begin{equation}
 \delta M=T\delta S+\Omega\delta J+\Phi\delta Q_e.
 \label{eq:first-law}
\end{equation}
All four parameters $(m,a,e,B)$ are varied in this relation.  The effects of
$B$ are contained in the variations of the physical charges and in the
solution-space dependence of the canonical generator.

\subsection{Integrability condition and zero-field boundary condition}

On any regular open set where the three one-forms
$\dd S$, $\dd J$, and $\dd Q_e$ are linearly independent, consider a
tangent vector $X$ in solution space that preserves all three extensive
variables,
\begin{equation}
 X[S]=X[J]=X[Q_e]=0 .
\end{equation}
The canonical first law then gives $X[M]=0$.  By the constant-rank theorem,
the mass therefore depends locally only on the three physical horizon
variables,
\begin{equation}
 M=M(S,J,Q_e).
 \label{eq:mass-state-function}
\end{equation}
Whenever $(S,J,Q_e,B)$ provides a valid local coordinate system, this is
equivalently expressed as
\begin{equation}
 \left(\frac{\partial M}{\partial B}\right)_{S,J,Q_e}=0 .
\end{equation}
The remaining function is fixed by the zero-field boundary condition.  At
$B=0$, the solution reduces to Kerr--Newman, with $J=ma$, $Q_e=e$, and
\begin{equation}
 S=\pi\left[(m+d_0)^2+a^2\right],
 \qquad d_0=\sqrt{m^2-a^2-e^2}.
 \label{eq:KN-entropy}
\end{equation}
For $m>0$, the map to $(S,J,Q_e)$ is locally invertible because
\begin{equation}
 \det\frac{\partial(S,J,Q_e)}{\partial(m,a,e)}
 =2\pi m\left(2m+d_0+\frac{m^2+a^2}{d_0}\right)>0.
 \label{eq:KN-Jacobian}
\end{equation}
Imposing
\begin{equation}
 M(m,a,e,0)=m
 \label{eq:zero-field-boundary}
\end{equation}
therefore fixes the mass on the complete Kerr--Newman state space.  In terms
of the extensive variables, the result is the positive branch of the
Christodoulou--Ruffini relation \cite{ChristodoulouRuffini},
\begin{align}
 M^2
 &=\frac{S}{4\pi}+\frac{Q_e^2}{2}
 +\frac{\pi(Q_e^4+4J^2)}{4S}
 \nonumber\\
 &=\frac{(R_H^2+Q_e^2)^2+4J^2}{4R_H^2},
 \qquad M>0.
 \label{eq:CR-mass}
\end{align}
The isolated-horizon construction assigns the same energy to an
axisymmetric Einstein--Maxwell horizon with area, angular momentum, and
charge $(A_H,J,Q_e)$ \cite{AshtekarRotatingIH},
\begin{equation}
 M_\Delta=
 \frac{\sqrt{(R_H^2+Q_e^2)^2+4J^2}}{2R_H}=M.
 \label{eq:isolated-horizon-mass}
\end{equation}
This agreement extends to KN--BR the compatibility between the canonical
integrability results and the isolated-horizon results found for
magnetized Kerr--Newman black holes \cite{AstorinoCompere,BoothMelvin}.

\section{Thermodynamics}
\label{sec:thermo}

\subsection{Properties of the canonical mass}

Equation~\eqref{eq:CR-mass} shows that the external BR field enters the
canonical mass through the physical horizon quantities rather than as an
additional argument of the fundamental relation.  In the original solution
parameters,
\begin{equation}
 M(m,a,e,B)=\frac12\left[
 R_H^2+\frac{2e^2}{K^2}
 +\frac{e^4+4m^2a^2}{K^4R_H^2}
 \right]^{1/2},
 \label{eq:mass-explicit}
\end{equation}
where
\begin{align}
 R_H^2&=\frac{r_+^2+a^2}
 {K[(1+kBr_+)^2+B^2r_+^2]},
 \nonumber\\
 r_+&=\frac{m+\mu}{1+k^2-e^2/a^2}.
 \label{eq:RH-explicit}
\end{align}
Together with Eq.~\eqref{eq:auxiliary-functions}, this gives the canonical
mass explicitly in $(m,a,e,B)$.

The mass is homogeneous in its extensive variables, and the first law
\eqref{eq:first-law} implies the Smarr formula
\cite{Smarr1973,TownsendLecture}
\begin{equation}
 M=2TS+2\Omega J+\Phi Q_e.
 \label{eq:Smarr}
\end{equation}
Holding $(m,a,e)$ fixed while changing $B$ gives
\begin{align}
 \left(\frac{\partial M}{\partial B}\right)_{m,a,e}
 =T\left(\frac{\partial S}{\partial B}\right)_{m,a,e}
 +\Omega\left(\frac{\partial J}{\partial B}\right)_{m,a,e}
 +\Phi\left(\frac{\partial Q_e}{\partial B}\right)_{m,a,e}.
 \label{eq:B-chain-rule}
\end{align}
This derivative is generally nonzero, even though
$(\partial M/\partial B)_{S,J,Q_e}=0$.  The external field changes the
map from the solution parameters to the physical charges, while the
fundamental relation among $M$, $S$, $J$, and $Q_e$ retains the
Kerr--Newman form.

\subsection{Thermodynamic potentials}

The potentials defined in Eq.~\eqref{eq:thermodynamic-definitions} are the
derivatives of the canonical mass,
\begin{align}
 T&=\left(\frac{\partial M}{\partial S}\right)_{J,Q_e}
 =\frac{1}{8\pi M}
 \left[1-\frac{Q_e^4+4J^2}{R_H^4}\right],
 \nonumber\\
 \Omega&=\left(\frac{\partial M}{\partial J}\right)_{S,Q_e}
 =\frac{J}{M R_H^2},
 \nonumber\\
 \Phi&=\left(\frac{\partial M}{\partial Q_e}\right)_{S,J}
 ~=\frac{Q_e(R_H^2+Q_e^2)}{2M R_H^2}.
 \label{eq:thermodynamic-potentials}
\end{align}
Comparison with Eq.~\eqref{eq:thermodynamic-definitions} determines the
canonical generator,
\begin{equation}
 \alpha=\frac{T}{T_H},
 \qquad \Omega_{\rm int}=\Omega_H-\frac{\Omega}{\alpha},
 \qquad \Phi_{\rm int}=\Phi_H-\frac{\Phi}{\alpha}.
 \label{eq:generator-coefficients}
\end{equation}
Substitution of the horizon quantities gives
\begin{align}
 \alpha&=\frac{K(R_H^4-Q_e^4-4J^2)}
 {4M\mu R_H^2},
 \nonumber\\
 \Omega_{\rm int}&=\frac{aK}{r_+^2+a^2}
 -\frac{4\mu J}{K(R_H^4-Q_e^4-4J^2)},
 \nonumber\\
 \Phi_{\rm int}&=\frac{er_+}{r_+^2+a^2}
 -\frac{2\mu Q_e(R_H^2+Q_e^2)}
 {K(R_H^4-Q_e^4-4J^2)}.
 \label{eq:generator-explicit}
\end{align}
Equations~\eqref{eq:physical-charges}, \eqref{eq:RH-explicit}, and
\eqref{eq:generator-explicit} express the canonical generator entirely in
terms of $(m,a,e,B)$.

At a regular extremal horizon,
\begin{equation}
 R_H^4=Q_e^4+4J^2,
 \qquad \mu=0.
 \label{eq:extremal-relation}
\end{equation}
The mass remains regular.  The explicit nonextremal ratios in
Eq.~\eqref{eq:generator-explicit} are defined at extremality by the
continuous limit of the chosen canonical generator, consistently with the
local extremal isolated-horizon relation
\cite{LewandowskiPawlowskiExtremal}.

\subsection{The canonical frame}

For each solution, the canonical symmetry parameter can be written as a pure
time translation after a change of coordinates and gauge
\cite{AstorinoCompere,ComperePForm}.  Define
\begin{equation}
 t_{\rm can}=\frac{t}{\alpha},
 \qquad \phi_{\rm can}=\phi-\Omega_{\rm int}t,
 \label{eq:canonical-coordinates}
\end{equation}
and perform the gauge transformation
\begin{equation}
 A^{\rm can}=A+\dd(\Phi_{\rm int}t).
 \label{eq:canonical-gauge}
\end{equation}
Then
\begin{equation}
 A^{\rm can}_{t_{\rm can}}
 =\alpha(A_t+\Omega_{\rm int}A_\phi+\Phi_{\rm int}),
 \qquad
 \epsilon_M=(\partial_{t_{\rm can}},0).
 \label{eq:canonical-frame}
\end{equation}
The mass is therefore the conserved charge associated with time translation
in the canonical frame and gauge.

\subsection{Alternative thermodynamics with the external field as a source}

The canonical construction varies all parameters, including $B$.  A
different thermodynamics results when $B$ is held fixed as an external
source.  In that case the integrability condition is imposed only within a
fixed-$B$ family and leaves one undetermined function, the alternative mass
$\widetilde M(S,J,Q_e;B)$, to be selected by an additional boundary
prescription \cite{AstorinoCompere}.  The corresponding infinitesimal
canonical charge can be written as
\begin{equation}
 \slasheddelta M
 =\delta\widetilde M+\mu_{\rm mag}\delta B
 =\widetilde T\delta S+\widetilde\Omega\delta J
 +\widetilde\Phi\delta Q_e,
 \label{eq:alternative-first-law}
\end{equation}
where
\begin{equation}
 \mu_{\rm mag}=-\left(
 \frac{\partial\widetilde M}{\partial B}
 \right)_{S,J,Q_e}.
 \label{eq:magnetic-moment}
\end{equation}
The quantity $\mu_{\rm mag}$ is the potential conjugate to the external
field, and the tilde potentials are the derivatives of $\widetilde M$ at
fixed $B$.  This fixed-source description concerns a different choice of mass
and boundary conditions from the canonical mass in
Eq.~\eqref{eq:CR-mass} \cite{GibbonsPangPope,KOPthermo,AstorinoCompere}.

\subsection{Kerr--BR limits}

The two previously studied Kerr--BR thermodynamics follow from the general
construction.  At $B=0$, $K=1$ and
\begin{equation}
 M=m,
 \qquad (\alpha,\Omega_{\rm int},\Phi_{\rm int})=(1,0,0).
 \label{eq:zero-field-limit}
\end{equation}
On the neutral branch $e=0$, Eq.~\eqref{eq:neutral-definitions} gives
\begin{align}
 Q_e&=0,
 \qquad J=\frac{ma}{K^2},
 \qquad M=\frac{m}{Kb},
 \nonumber\\
 \alpha&=\frac1b,
 \qquad \Omega_{\rm int}=0,
 \qquad \Phi_{\rm int}=0.
 \label{eq:neutral-limit}
\end{align}
These are the genuine Kerr--BR$_0$ quantities of Ref.~\cite{KOPthermo}.
The factor $\beta=b$ used there to rescale the timelike Killing vector is
related to the present coefficient by $\alpha=\beta^{-1}$.

The specially charged branch is defined by
\begin{equation}
 e=e_s\equiv\frac{maB}{\sqrt{1-a^2B^2}},
 \label{eq:special-charge}
\end{equation}
for which the general expressions reduce to the Kerr--BR$_s$
thermodynamics of Ref.~\cite{HuCaiWang}. Since our conventions differ slightly from those used in \cite{HuCaiWang}, we show the mappings between the parameters in two works in Appendix~\ref{app:limits}.

The canonical mass is a quasi-local horizon energy and an integrable charge
on the KN--BR solution space.  An energy assigned to an outer boundary also
depends on the choice of time translation, electromagnetic boundary data,
and reference prescription there \cite{GibbonsPangPope,
AstorinoCompere}.  The two notions correspond to different Hamiltonian
boundary-value problems.

\section{Conclusions}
\label{sec:conclusions}

In this work, we have extended the thermodynamics of the neutral and specially charged
Kerr--BR black holes to the black holes in general KN--BR family. The electric and
external-field parameters are independent in this family, so the canonical
integrability conditions can be imposed on the full four-parameter solution
space with the complete Kerr--Newman family as the zero-field boundary.

The physical angular momentum follows from an analytic horizon integral
that combines its gravitational and electromagnetic contributions.  The
infinitesimal charge associated with $(\partial_t,0)$ is conserved under
deformations of the integration surface but is nonintegrable in solution
space.  A general canonical generator restores integrability, and the
zero-field boundary condition determines the canonical mass and its
thermodynamic potentials.  The resulting mass obeys the
Christodoulou--Ruffini relation and agrees with the corresponding
isolated-horizon energy.

The Bertotti--Robinson field changes the map between the solution parameters
and the physical horizon charges, while the fundamental relation, first law,
and Smarr formula retain the Kerr--Newman form.  The general construction
contains the two known Kerr--BR thermodynamics as limits.  Treating the
external field as a fixed source instead leads to an alternative mass and a
magnetic work term determined by the chosen boundary prescription.

\begin{acknowledgments}
We thank Li Hu for useful discussions.  The work is partly supported by
NSFC Grant No. 12275004, 12547132, 12547127, and No. 12588101.
\end{acknowledgments}

\appendix

\section{Regular Maxwell potential and analytic horizon charges}
\label{app:charges}

In this appendix, we present the details of the horizon-charge computation.
Let $\hat\phi$ denote the angular coordinate of Ref.~\cite{OPnew}, whose
period is $2\pi/K$ and which is related to the coordinate used in the main
text by $\phi=K\hat\phi$.  With the arbitrary closed one-form omitted, the
Maxwell potential has components
\begin{align}
 A_t&=\frac{a}{\Omegac}\left[-(1+k^2)B
 +(mB\cos^2\theta-k)\frac{r}{\rho^2}\right],
 \nonumber\\
 A_{\hat\phi}^{\rm bare}
 &=\frac{1}{\Omegac}\bigg[(1+k^2)a^2B-kr-\frac1B
 \nonumber\\
 &\hspace{0.8cm}+\frac{r}{\rho^2}
 \left(-mB(r^2+a^2)\cos^2\theta
 +ka^2\sin^2\theta\right)\bigg].
 \label{eq:potential-bare}
\end{align}
The auxiliary quantities satisfy
\begin{align}
 u^2=1-(1+k^2)a^2B^2,\hspace{3ex}
 k+mB=\frac{eu}{a}, \hspace{3ex}
 \left.\Omegac^2\right|_{\rm axis}
 =\left(u+\frac{eBr}{a}\right)^2.
 \label{eq:axis-identities}
\end{align}
Since $\dd\hat\phi=\dd\phi/K$, the potential denoted by $A^{(0)}$ in the
main text is
$A^{(0)}=A_t\dd t+(A_{\hat\phi}^{\rm bare}/K)\dd\phi$.  On the smooth branch
with $\Omegac|_{\rm axis}=u+eBr/a>0$, one has
$A_{\hat\phi}^{\rm bare}|_{\rm axis}=-u/B$, and the shift in
Eq.~\eqref{eq:axis-gauge} makes the potential regular at both poles.  It also
gives
\begin{equation}
 A\longrightarrow-\frac{er}{\rho^2}
 \left(\dd t-a\sin^2\theta\dd\phi\right)
 \qquad (B\to0).
 \label{eq:KN-potential-limit}
\end{equation}

For the horizon integrals, set $r=r_+$ and define
\begin{align}
 x&=\cos\theta, &R&=r^2+a^2,
 &h&=r^2+a^2x^2,
 \nonumber\\
 H&=\sqrt{I_+}, &W&=mBr^2+ka^2.
 \label{eq:integral-notation}
\end{align}
The horizon equation and Eq.~\eqref{eq:axis-identities} imply
\begin{align}
 \mu&=m-\frac{(1+k^2)a^2}{r},\hspace{3ex}
 H=u+\frac{eBr}{a},
 \nonumber\\
 \mathscr R&\equiv r^2(k+mB)^2
 -\left[(1+k^2)R-2mr\right]
 \times\left[1-(1+k^2)a^2B^2\right]=0.
 \nonumber
\end{align}
Introduce
\begin{align}
 \mathcal E(x)&=\frac{aR}{H}
 \frac{x\left[k+(1+k^2)Br-B\mu x^2\right]}{h},
 \nonumber\\
 \mathcal A(x)&=\frac{rW}{H}\frac{1-x^2}{h},
 \nonumber\\
 \mathcal G(x)&=\frac{2aR(1-x^2)}{h}
 \left[\mu+\frac{rR(1+B^2\mu^2x^2)}{H^2h}\right].
 \label{eq:horizon-density-functions}
\end{align}
Direct differentiation gives
\begin{align}
 (\star F)_{\theta\phi}
 =\frac{\sin\theta}{K}\mathcal E'(x),\hspace{3ex}
 A_\phi=\frac{1}{K}\mathcal A(x),\hspace{3ex}
 (\star\dd\eta^\flat)_{\theta\phi}
 =\frac{\sin\theta}{K^2}\mathcal G(x),
 \nonumber
\end{align}
where the prime denotes $\dd/\dd x$.  At the two axes,
\begin{equation}
 \mathcal E(1)=\frac{a}{H}
 \left[k+(1+k^2)Br-B\mu\right]=e,\qquad
 \mathcal E(-1)=-e.
 \label{eq:E-endpoints}
\end{equation}
The electric flux is therefore
\begin{equation}
 Q_e=\frac{1}{2K}\left[\mathcal E\right]_{-1}^{1}
 =\frac{e}{K}.
 \label{eq:Q-integral}
\end{equation}
The same identities reduce the horizon potential to
$\Phi_H=er/(r^2+a^2)$.

The total angular-momentum density contains the combination
$\mathcal G+4\mathcal A\mathcal E'$, an antiderivative of which is
\begin{equation}
 \mathcal Y(x)=\frac{x(v_0+v_2x^2+v_4x^4)}{H^2h^2},
 \label{eq:Y-primitive}
\end{equation}
where
\begin{align}
 v_0&=2aR\left[\mu H^2r^2+rR
 +2rW\left(k+(1+k^2)Br\right)\right],
 \nonumber\\
 v_4&=2aR\mu\left[-a^2H^2-rRB^2\mu+2rBW\right],
 \nonumber\\
 v_2&=4maH^2R^2-v_0-v_4.
 \label{eq:Y-coefficients}
\end{align}
Using $H^2=1+2kBr+(1+k^2)B^2r^2$, one finds
\begin{equation}
 \mathcal Y'-\mathcal G-4\mathcal A\mathcal E'
 =-\frac{4ax^2R^2(a^2x^2-r^2)}{rH^2h^3}\,\mathscr R=0.
 \label{eq:Y-identity}
\end{equation}
Since $\mathcal Y(\pm1)=\pm4ma$, the total angular momentum is
\begin{equation}
 J=\frac{1}{8K^2}\int_{-1}^{1}
 \left(\mathcal G+4\mathcal A\mathcal E'\right)\dd x
 =\frac{1}{8K^2}\left[\mathcal Y\right]_{-1}^{1}
 =\frac{ma}{K^2}.
 \label{eq:J-integral}
\end{equation}
For generic charged KN--BR configurations, both the gravitational and Maxwell
terms contribute. 

\section{Relations to the Kerr--BR subfamilies}
\label{app:limits}

The specially charged Kerr--BR$_s$ branch is specified by
Eq.~\eqref{eq:special-charge}, for which $k=0$. At first glance, it seems that our results cannot reduce to those of \cite{HuCaiWang}. The discrepancy originates from different conventions for Maxwell potentials.
To compare with the
parameters of Ref.~\cite{HuCaiWang}, use the map of
Ref.~\cite{OPnew}
\begin{align}
 I_1=1-\frac12a^2B^2,\hspace{3ex}
 I_2=1-a^2B^2,\hspace{3ex}
 m_{\rm HCW}=\widetilde m=m\frac{I_1}{I_2}.
 \nonumber
\end{align}
The conicity factor becomes
\begin{equation}
 K=P_0=1+B^2\left(\frac{\widetilde m^{\,2}I_2}{I_1^2}-a^2\right),
 \qquad J=\frac{ma}{P_0^2}.
 \label{eq:P0-map}
\end{equation}
Together with the entropy and electric charge, these relations reduce the
mass and generator to the expressions of Ref.~\cite{HuCaiWang}. 

At the same value of $B$, the two metric conventions agree while their
axis-regular Maxwell potentials have opposite orientations,
\begin{equation}
 A_{\rm HCW}(B)=-A_{\rm KNBR}(B).
 \label{eq:potential-sign-map}
\end{equation}
Consequently,
\begin{equation}
 (Q_e,\Phi_H,\Phi_{\rm int},\Phi)_{\rm HCW}
 =-(Q_e,\Phi_H,\Phi_{\rm int},\Phi)_{\rm KNBR},
 \label{eq:electric-sign-map}
\end{equation}
whereas $M$, $J$, $\alpha$, $\Omega_H$, and $\Omega_{\rm int}$ coincide.
The electric work term is unchanged because both the charge and its conjugate
potential reverse.  In particular,
\begin{equation}
 \Phi_{\rm int}^{\rm KNBR}
 =-\Phi_{\rm int}^{\rm HCW}
 =-\frac{4B^3\widetilde m^{\,2}a}
 {(2-B^2a^2)^2\sqrt{1-B^2a^2}}.
 \label{eq:HCW-Phi-int}
\end{equation}
Equivalently, identifying $B_{\rm HCW}=-B_{\rm KNBR}$ aligns the physical
external-field direction and removes the additional electric sign reversal.

\bibliography{refs}
\end{document}